# Colloidal Anisotropic ZnO-Fe@Fe$_x$O$_y$ Nanoarchitectures with Interface-Mediated Exchange-Bias and Band-Edge Ultraviolet Fluorescence

Athanasia Kostopoulou,[1,2] Franck Thétiot,[1] Ioannis Tsiaoussis,[3] Maria Androulidaki,[1] P. Davide Cozzoli,[4,5] and Alexandros Lappas[1,*]

[1]Institute of Electronic Structure and Laser, Foundation for Research and Technology-Hellas, Vassilika Vouton, 71110 Heraklion, Greece

[2]Department of Chemistry, University of Crete, Voutes, 71003 Heraklion, Greece

[3]Department of Physics, Aristotle University of Thessaloniki, 54124 Thessaloniki, Greece

[4]National Nanotechnology Laboratory (NNL), Istituto Nanoscienze- CNR, c/o Distretto Tecnologico ISUFI, via per Arnesano km 5, 73100 Lecce, Italy.

[5]Dipartimento di Matematica e Fisica "E. De Giorgi", Università del Salento, via per Arnesano, 73100 Lecce, Italy.

Supporting Information

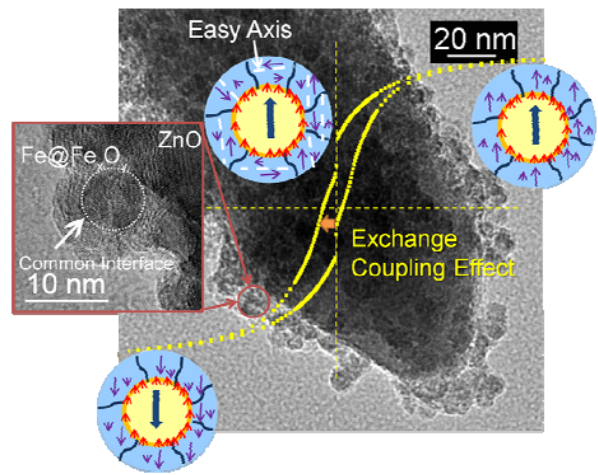

**ABSTRACT**: Hybrid nanocrystals (HNCs), based on ZnO nanorods (NRs) decorated with magnetic Fe-based domains, were synthesized via a colloidal seeded-growth method. The approach involved heterogeneous nucleation of Fe nanocrystals on size-tailored ZnO nanorod seeds in a noncoordinating solvent, followed by partial surface oxidation of the former to the corresponding Fe@Fe$_x$O$_y$ core@shell domains. HNCs with variable population and size of the Fe-based nanodomains could be synthesized depending on the surface reactivity of the ZnO seeds. The structure-property relationships in these HNCs were carefully studied. In HNCs characterized by a large number of small Fe@Fe$_x$O$_y$ core@shell nanodomains on the ZnO seed surface, the interfacial communication across the Fe-core and Fe$_x$O$_y$-shell generated a sizeable exchange-bias effect mediated by frozen interfacial spins. On the other hand, in HNCs carrying a lower density of comparatively larger Fe@Fe$_x$O$_y$ domains, partial removal of the Fe core created an inner void in-between that led to suppressed exchange coupling anisotropy. As a further proof of functionality, the HNCs exhibited pronounced band-edge ultraviolet fluorescence. The latter was blue-shifted compared to the parent ZnO NRs, inferring coupling of the semiconductor and magnet sections.

## 1. INTRODUCTION

The growing request for advanced nanoscale entities capable to address multiple technological tasks and/or exhibit unconventional physical-chemical behavior is stimulating intense efforts towards development of novel multi-component nanostructures, in which non-homologous properties of different materials are allowed to coexist, synergistically reinforce and/or exchange couple with each other.[1, 2] Recently, the versatile armory of colloidal chemistry synthetic tools to design and grow complex nanostructures has paved the way to several prototypes of all-inorganic hybrid nanocrystals (HNCs).[1, 3-7] HNCs are heterostructured nanoparticles made of domains of diverse chemical nature, structure, and geometry, which are welded together through direct bonding junctions without any intermediate bridging molecule. They can engage semiconductor or/and metal compounds assembled in various topologies, ranging from concentric core@shell configurations,[1, 8-16] to non-onion-like hetero-oligomers with peanut-, dumbbell-, or branch-type connectivity based on associations of spherical,[6, 12, 17-20] and/or anisotropically shaped material sections.[1, 3-5, 21, 22]

The coexistence, within the same nanocrystal entity, of distinct material sections directly interconnected through bonding interfaces not only enables multifunctionality due to simultaneous availability of complementary properties, but also establishes the potential for exchange coupling mechanisms between different (e.g. magnetic and optical) properties via the electronic contact junctions holding across interfaced material domains. Such novel nanomaterials have been already implemented in various applications in fields, such as catalysis[18, 21, 23-25], biology/nanomedicine[19, 26-29] and optoelectronics[30-32].

Although a broad variety of colloidal nanoheterostructures has been developed for disparate combinations of materials, HNCs involving coupled magnetic and semiconductor compounds are comparatively more limited. Most prominent examples pertain to HNCs incorporating fluorescent metal chalcogenide semiconductors, such as $\gamma$-Fe$_2$O$_3$-MeX (with Me= Zn, Cd, Hg and X= S, Se)[19, 27], Fe$_3$O$_4$-CdS[7], FePt@CdS(Se) heterodimers, ternary PbX (X= S, Se)-Au-Fe$_3$O$_4$ heteroligomers [13, 18], Co@CdSe core@shell nanospheres[33] and Co-tipped CdSe@CdS core@shell nanorods.[5] On the side of oxide semiconductors, though, fewer cases of prototype hybrid systems have been demonstrated,[2] such as, for example, HNCs based on phase- and shape-tailored anisotropic TiO$_2$ nanocrystals with facet-dependent chemical reactivity, which have been decorated with spherical $\gamma$-Fe$_2$O$_3$[3, 21, 34] or Co[22] domains epitaxially grown at selected locations. Such bi-functional nanoarchitectures are expected to open new opportunities in environmental and energy applications, serving, for instance, as magnetically recoverable photocatalysts, in biomedicine as platforms for analytical separation, and as multitask entities for drug delivery and multimodal imaging, as well as in spintronics, as possible sources of spin-polarized current and/or light.[2, 35]

Among oxide semiconductors though, zinc oxide (ZnO) has drawn much attention over the last decade due to its direct wide band gap ($E_g$= 3.37eV) and high exciton binding energy (60 meV). Most technological potential of this material at the nanoscale is enabled by its tunable optoelectronic properties, including the intrinsically size-dependent exciton absorption edge and the related ultraviolet band-edge fluorescence, on one side, and the visible green emission due to intrinsic or extrinsic deep impurity levels, on the other side. In that respect novel light emitting devices may be developed by growing ZnO-based composite materials in various structures and especially when low-spatial dimensionality is involved. Confinement effects can then enhance exciton oscillator strength and provide improved quantum efficiency in tailored-made ZnO nanostructures, raising the technological potential of such materials. Nevertheless, there have been only a few literature reports on ZnO-based nanoheterostructures with controlled topologies that go beyond conventional composite systems made of randomly arranged material particles. Examples include: HNCs where spherical Fe$_3$O$_4$ or FePt cores are surrounded by a conformal ZnO shell[15, 36], which show superparamagnetic behaviour and weak UV emission; in another account, Fe$_3$O$_4$ nanorod cores were protected by a polycrystalline ZnO shell, leading to a binary system exhibiting microwave-absorption properties[37]. In another approach, a hydrothermal method was used for the growth of ZnO microrod templates and the subsequent preparation of ZnO/iron oxide heterostructured composites, with magnetic and optical properties[38]. The present work makes an advance in this field, as it reports for the first time on the colloidal synthesis of anisotropic HNCs with a ternary phase composition, which individually consist of ZnO nanorods uniformly covered with size-tunable concentric core@shell nanoparticles made of a Fe-metal nanocrystal core passivated by an Fe$_x$O$_y$ polycrystalline shell. Henceforth, these are referred to as ZnO-Fe@Fe$_x$O$_y$ HNCs. We demonstrate that distinct nanosized ZnO and Fe$_x$O$_y$ sections directly connected through bonding interfaces enable multifunctionality pertaining to band-edge UV fluorescence as well as weak ferromagnetic behaviour at room temperature. Their dual physical response benefits their technological potential, straddling magnetic storage, optoelectronics and even biomedicine.

## 2. EXPERIMENTAL SECTION

**2.1 Materials.** All reagents were of relatively high-purity and were used as received without further purification. Anhydrous zinc acetate (Zn(CH$_3$COO)$_2$, 99.99%, Zn(Ac)$_2$) was purchased from Riedel. Iron pentacarbonyl (Fe(CO)$_5$, 99.999%) and oleylamine (CH$_3$(CH$_2$)$_7$CH=CH(CH$_2$)$_7$CH$_2$NH$_2$, 70%, OLAM) were purchased from Aldrich, whereas 1-hexadecylamine (CH$_3$(CH$_2$)$_{14}$CH$_2$NH$_2$, 90%, HDA) and 1-octadecene (ODE, CH$_3$(CH$_2$)$_{15}$CH=CH$_2$, 90%, ODE) were purchased from Alfa Aesar. All solvents were of analytical grade. Chloroform (CHCl$_3$) and absolute ethanol were purchased from Aldrich, while 2-propanol was purchased from Riedel.

**2.2 HNC synthesis.** All syntheses were carried out under argon atmosphere in 50-mL round-bottom three-neck flasks connected via a reflux condenser to standard Schlenk line setup, equipped with immersion temperature probes and digitally-controlled heating mantles. The air-/moisture- sensitive precursors, namely Fe(CO)$_5$ and Zn(Ac)$_2$ were stored and handled under argon atmosphere in a glove box (MBRAUN, UNILab). The general approach to synthesize the ZnO-Fe@Fe$_x$O$_y$ HNCs relied on a two step seeded-growth scheme, involving high temperature heterogeneous nucleation of multiple Fe domains onto preformed ZnO nanorods (NRs), followed by partial oxidation of the former.

*a) Synthesis of the ZnO NRs seeds.* The ZnO NRs that were used as seeds were synthesized by a modified scaled-up literature protocol.[39] Two batches of ZnO NRs samples were prepared with a similar procedure. To prepare the first batch of NRs sample (NR-1), 6.00 g (22 mmol) of HDA was carefully degassed in a flask under vacuum at 100 °C for 30 min, after which the mixture was cooled to 80 °C and kept under argon flow. Then 2.05 g (11 mmol) of Zn(Ac)$_2$ (the HDA to Zn(Ac)$_2$ molar ratio was 2:1) was quickly transferred from an argon-filled sealed vial to the flask preheated at 80 °C under vigorous stirring. The resulting mixture was quickly heated to 240 °C (at 20 °C/min) and annealed at this temperature for 10 min. During this process, the initially clear yellow solution became white turbid. At the end of the heating, the mixture was cooled down to 80 °C and subsequent extraction/purification procedures were performed under ambient atmosphere. 2-propanol was added to induce flocculation of the ZnO product. The precipitated ZnO was collected by centrifugation at 6000 rpm, washed three times with absolute ethanol and finally re-dispersed in CHCl$_3$. Further purification to remove precursor and residual surfactants was accomplished by performing three cycles of centrifugation at 6000 rpm for 15 min and redispersion in CHCl$_3$.

For the second batch of NRs seeds sample (NR-2), the same quantity of the HDA and 0.83 gr (3.1 mmoles) of OLAM (the HDA:OLAM:Zn(Ac)$_2$ molar ratio was 2:0.3:1) were heated up to 280 °C for the same time. All the other synthesis parameters were kept the same.

*b) Seeded-growth synthesis of HNCs.* In a typical synthesis of HNCs, 41mg (0.5 mmol) of already purified ZnO NRs with a selected size were dispersed in 20 mL of ODE under vigorous stirring. The resulting mixture was degassed for 30 min at 120 °C and then heated up to 235° C under argon flow. Subsequently, 1 mL of a 0.5 M Fe(CO)$_5$ solution in previously degassed ODE was quickly added to the flask in a single shot via a disposable syringe. The fast iron precursor injection induced a sudden drop of the reaction mixture temperature (by 20-40 °C), after which the temperature slowly recovered to the initial value. Release of white vapors was also observed after the Fe(CO)$_5$ addition and the solution turned from milky to dark brown in a few minutes, indicating the Fe(CO)$_5$ decomposition and nucleation of the metallic Fe. The mixture was annealed for 1 h to allow completion of growth of the iron NCs onto the preformed ZnO NRs seeds, as well as formation of their Fe$_x$O$_y$ shell. The mixture was then cooled down to 120 °C and exposed to air for 1 h to induce further oxidation of the initially formed Fe-component. Then the reaction was stopped by removing the heating mantle. The HNCs were precipitated upon 2-propanol addition to the crude mixture at room temperature, separated by centrifugation at 6000 rpm, washed six times with 2-propanol, and finally re-dispersed in CHCl$_3$.

## 2.3 Characterization techniques

*a) Transmission electron microscopy (TEM).* Low-magnification TEM and phase-contrast high-resolution TEM (HRTEM) images were recorded on a JEOL 2100 and a JEOL 2011 (with an atomic resolution of 0.194 nm) transmission electron microscopes, operating at an accelerating voltage of 200 kV. For the purposes of the TEM analysis, a drop of a diluted colloidal nanocrystal solution in CHCl$_3$ was deposited onto a carbon-coated copper TEM grid and then the solvent was allowed to evaporate. Statistical analysis was carried out on several wide-field low-magnification TEM images, with the help of dedicated software (Gatan Digital Micrograph). For each sample, about 150 individual particles were counted up. All the images were recorded by the Gatan ORIUS™ SC 1000 CCD camera and the structural features of the nanostructures were studied by two-dimensional (2D) fast Fourier transform (FFT) analysis.

*b) X-ray Diffraction (XRD).* Powder X-ray diffraction (XRD) studies were performed on a Rigaku D/MAX-2000H rotating anode diffractometer with Cu-K$_α$ radiation, equipped with a secondary graphite monochromator. The XRD data at room temperature were collected over a 2θ scattering range of 5-90°, with a step of 0.02° and a counting time of 10 sec per step.

*c) Magnetic measurements.* The magnetic properties of the samples were studied by a Superconducting Quantum Interference Device (SQUID) magnetometer (Quantum Design MPMS-XL5). Hysteresis loops of the magnetization, M(H), were obtained at 5 K after cooling the powder samples either in a zero field (ZFC) or at an applied field H= 10 kOe (FC); then the field was swept from +10 kOe to -10 kOe and back again to the maximum positive field value. Additional, magnetization data against temperature, under ZFC and FC conditions, were recorded at a 10 kOe applied field and in the range between 5 and 300 K. In all such measurements the samples were first maintained at 300 K for about 10 min, at a zero-field and then cooled at a rate of 6 K/min to the base temperature.

*d) Photoluminescence (PL) experiments.* Photoluminescence (PL) experiments were performed on solid-state samples produced after drying a small volume of the corresponding colloidal solution on a piece (10×10 mm$^2$) of a single-crystalline silicon <100> wafer. For the sample excitation we used a He-Cd CW laser operating at a wavelength of 325 nm, with 35 mW power. The PL spectra were measured at 300 K and resolved by using a UV grating, with 600 grooves/mm and a sensitive, calibrated liquid nitrogen-cooled CCD camera.

## 3. RESULTS AND DISCUSSION

**3.1 Growth of size-tailored core@shell Fe@Fe$_x$O$_y$ nanodomains on the ZnO NRs.** We have devised a colloidal chemistry pathway for the synthesis of anisotropic HNCs that are individually comprised of a single rod-shaped ZnO section, ubiquitously decorated with multiple nearly spherical Fe@Fe$_x$O$_y$ domains. Our synthetic protocol relies on a three-step seeded-growth approach[1] that involves: (i) the independent high-temperature synthesis, extraction and purification of amine-capped ZnO NRs; (ii) their subsequent exploitation as seed substrates for the heterogeneous nucleation and growth of small Fe nanocrystals upon pyrolysis of calibrated Fe(CO)$_5$ amounts in surfactant-free non-coordinating ODE; (iii) sacrificial surface oxidation of the Fe by oxygen upon air-exposure, generating a passivating Fe$_x$O$_y$ layer around each of the original Fe nanocrystals.[40, 41]

*ZnO nanorod structures.* In more detail, monodisperse ZnO NRs seeds were grown by aminolytic decomposition of Zn(Ac)$_2$ in an alkyl amine environment at 240 °C (NR-1, Figure S1a) and at 280 °C (NR-2, Figure S1b). In this procedure the amine molecules play two complementary roles: first, they operate as nucleophilic agents that attack the carbon atom of the carbonyl moiety in Zn(Ac)$_2$, thereby leading to the generation of the actual ZnO monomers (zinc hydroxo/oxo molecular or radical species), which will eventually condense and build the oxide lattice;[39, 42, 43] and second, they can act as facet-preferential coordinating agents for the resulting nanocrystals, regulating their size and shape evolution.[35] Previous studies have shown that reactions carried out at relatively high-er amine to zinc precursor molar ratios, in conjunction with high decomposition temperature, result in NRs with larger diameter (D) but lower aspect ratios (L/D).[39] We carried out the synthesis by utilizing both mono- and bi- surfactant mixtures. In HDA liquid media, at a HDA to Zn(Ac)$_2$ molar ratio of 2:1 and at 240 °C, bullet-like NRs were grown with a distinguished arrow-shaped profile at one apex and a flat termination at the opposite side. These nanostructures possessed an aspect ratio of ~3.4, with mean diameter and length dimensions of $D_1$= 70.4±11.5 nm, $L_1$= 239.3±30.5 nm, respectively (Figure S1a, c, d). In HDA:OLAM mixture, with a higher total amine to Zn(Ac)$_2$ molar ratio of 2.3:1 and at 280 °C, similarly shaped NRs with larger volume were generated. These were characterized by $D_2$= 83.4±13.3 nm, $L_2$= 266.8±37.3 nm and an aspect ratio of ~3.2 (Figure S1b, e, f). Such OLAM-driven size modulation should correlate with the particular steric hindrance posed by unsaturated alkyl chain structure of OLAM, which could affect the ultimate protecting and stabilizing capability of the mixed acetate/amine capping layer that dynamically adhered to the nanocrystal surface during growth. These results are in agreement with the evolution of the geometric features of ZnO nanostructures generated in similar alkyl amine media.[39] However, a significantly higher degree of size/shape homogeneity has clearly been achieved in the present case upon careful adjustment of the synthetic procedure and optimization of the relevant process parameters achieved in the present case. The NRs were found to be single-crystalline wurtzite ZnO (a=b=3.25±0.02 Å, c= 5.20±0.02 Å), with a preferred growth direction along the [0001] of the hexagonal structure (Figure 1a-c). This was clearly shown through the FFT analysis of the relevant HRTEM images (inset, Figure 1c), which allowed identification of characteristic lattice fringes; for example, in Figure 1c the spacing of 0.26 nm could be indexed well to the (0002) planes. It is known that the polar (000±1) planes of ZnO nanocrystals are generally distinguished by higher surface energy, compared to other facets (such as those in the perpendicular <01-10> directions) due to their inherent atomic structure and the lower degree of ligand passivation attained thereon. This accounts for the observed anisotropic development regime that ultimately leads to NRs elongated in the c-axis direction.[39, 44, 45] In particular, the bullet-like morphology originates from strongly unidirectional growth along the c-axis, a process that can indeed be promoted by the intrinsic structural (hence chemical) dissimilarity of the top and bottom basal (000±1) sides of c-axis elongated wurtzite lattice. Because of such condition, the less stable oxygen-rich (0001)-type facet, which corresponds to the fastest-growing direction, tends to disappear in favor of other (more stable) oblique facets, thereby evolving into an arrowhead apex.[1, 3-5, 46, 47]

*Fe-based nanocrystal structures.* Injection of a controlled amount of Fe(CO)$_5$ into a hot dispersion of the purified ZnO NRs in surfactant-free ODE, followed by air-exposure, led to HNCs that were individually composed of one of the original ZnO seeds decorated with a variable number of nearly spherical Fe-based nanoparticles. Two representative examples are shown in Figures 1d-i, henceforth referred to as *HNC-1* and *HNC-2*, which are distinguished by a relative high and low surface coverage of the relevant ZnO NR cores, respectively. Under the chosen syntheses conditions Fe(CO)$_5$ decomposition resulted in the direct deposition and growth of Fe-containing domains onto the ZnO seeds, rather than in the independent generation of free-standing nanoparticles throughout the bulk solution. This outcome can be rationalized within the frame of the classical nucleation theory,[48] which predicts that the activation energy required for heterogeneous nucleation of secondary material clusters onto pre-existing seeds and their subsequent enlargement is considerably lower than the energy barrier that has to be overcome for inducing the homogeneous nucleation of separate embryos of the same material.[1, 2, 35] Upon careful inspection of the TEM micrographs of the HNCs (Fig. 1e,f and Fig. 1h,i), the newly formed nanodomains covering the ZnO NRs were recognized to exhibit a much darker image contrast in their center area, while the thick edges around such regions appeared lighter. This contrast difference is more accentuated in the low-magnification images of the larger nanodomains attached on the surface of *HNC-2* (Fig. 1h) and less obvious in the case of the *HNC-1* (Fig. 1e). The observed lighter contrast in the latter postulates that a minority of hollow or fully oxidized nanodomains is likely present on the ZnO NR surface. Careful observation, though, of the corresponding HRTEM images near the semiconductor surface corroborates the coexistence of the afore-mentioned bimodal contrast features (Fig. 1f). Taking into account the known decomposition paths of the Fe(CO)$_5$ and the susceptibility of nanoscale metallic Fe to oxidation in hot mixtures with O$_2$ traces[40, 41] or exposed to air,[49, 50] the darker cores could preliminarily be interpreted as being metallic Fe due to its higher electron density and electron diffracting powder, while the outer shell could be made of an Fe$_x$O$_y$ phase.

Further evidence for a core@shell-type topological arrangement in the decorating nanodomains was provided by the FFT calculated for the selected areas in the HRTEM images, as indicated in Figures 1f,i. Analysis of the FFT patterns allows identification of two sets of coexisting diffraction spots (in addition to those associated with the underlying ZnO structure), which correspond to the two main phase components of the Fe-based domains grown on the ZnO seeds (inset, Figures 1f, i). One set can be indexed well to the (110) family of lattice planes of the α-Fe located in the core, while the second set can be attributed to the (311) family of lattice planes of the inverse cubic spinel Fe$_x$O$_y$ (of indistinguishable Fe$_3$O$_4$ and/or γ-Fe$_2$O$_3$ nature) in the outer shell section. The oxide shell was composed of small crystallites (<5 nm size) randomly oriented relative to one another (Figure S2).[40] Its good crystallinity most likely arose from the high temperature (235 °C) at which the shell formation occurred, as demonstrated by the fact that HNCs grown at lower temperatures (215 °C) accommodated similar core@shell type nanoparticles that, however, possessed a quasi-amorphous Fe$_x$O$_y$ shell (Figure S3 a-b). We believe that the annealing period at 235 °C during the synthesis of the HNCs allowed for the growth of the α-Fe NCs that had nucleated upon Fe(CO)$_5$ pyrolysis, while it also initiated their partial oxidation by O$_2$ traces present in the reaction liquid environment, leading to formation of a crystalline Fe$_x$O$_y$ shell. The oxidation was completed and the growth of the passivating oxide shell was terminated during the subsequent air-exposure step at lower temperature (120 °C).

*Morphological evolution of the HNCs.* Time-dependent TEM monitoring of the HNC growth indicated that no full oxidation of the Fe-based domains occurred for both *HNC-1* and *HNC-2*.

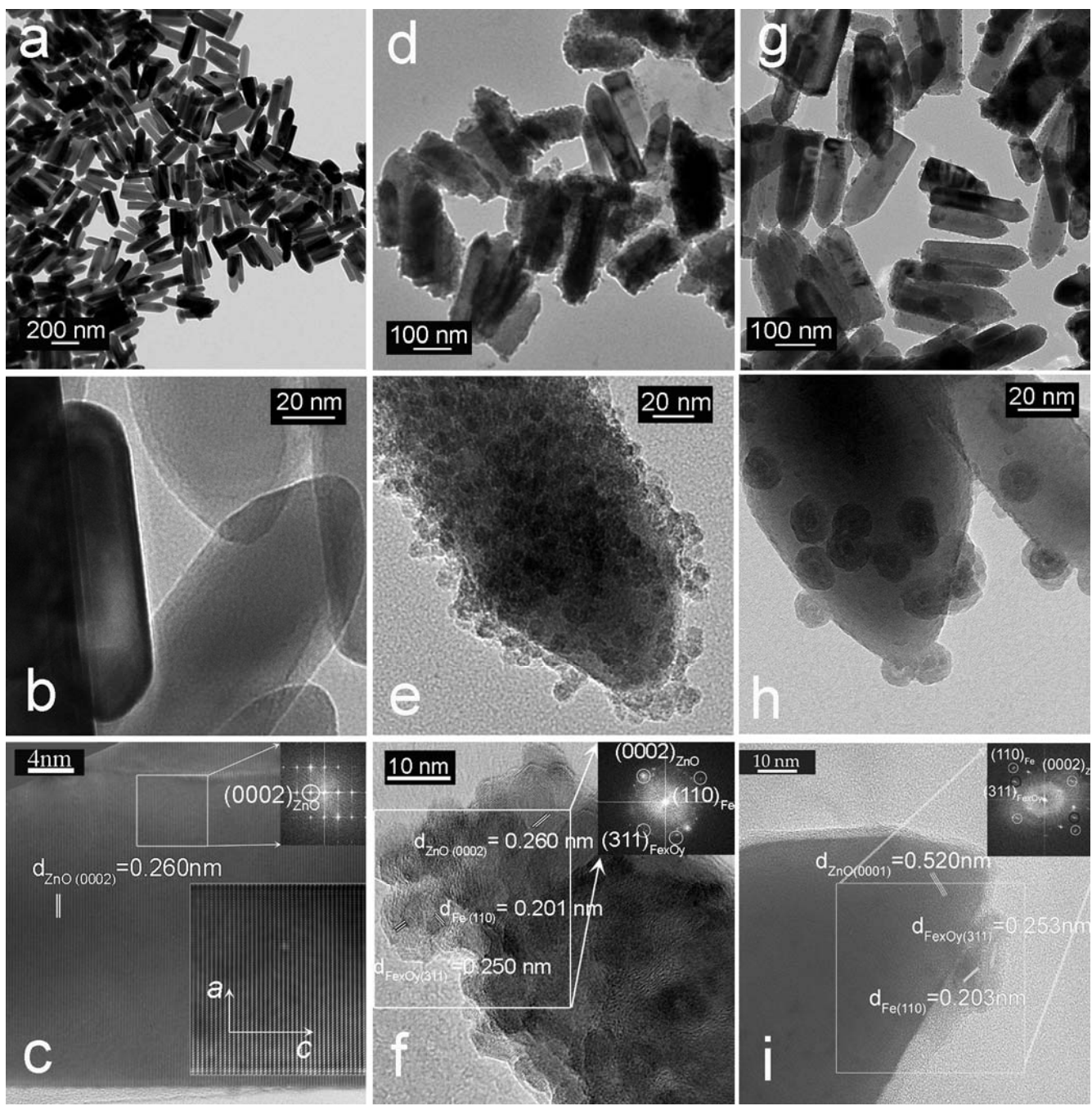

**Figure 1**. Representative low-magnification TEM (a, b) and HRTEM images (c) of the ZnO NRs seeds (sample NR-1) and of HNCs thereof with high (d, e, f) and low (g, h, i) degree of Fe@Fe$_x$O$_y$ domain surface coverage,, respectively (denoted as *HNC-1* and *HNC-2*, in the text). Insets: The corresponding FFT patterns calculated for the selected area marked by the rectangle in panels c, f, i. The zone axis in the images of panels c-,f,i, is the [01-10].

The outermost exposed layers of the metal domain underwent fast conversion into a polycrystalline oxide shell that appeared to prevent the core from further oxidation. However, an additional morphological evolution over time was recognized for the *HNC-2* decorated with the larger nanoparticle domains (Figure 1h). Our HRTEM studies revealed the formation of a void at least ~1 nm thick, which intervened between the core and the shell in the decorating nanodomains. Such features are reminiscent of a known hollowing process that can cause the selective etching of the interior of a nanostructure, commonly attributed to the so-called Kirkendall effect.[1, 2, 49, 50] The underling process was previously shown to involve a faster migration of the Fe atoms from the core to the outer shell than that of the oxygen atoms inwards, through a microscopic mechanism entailing vacancy coalescence and electron transport tunneling.[1, 2, 51] The latter effect was found to be operative only for shells with thicknesses of up to 1-3 nm and only metal nanoparticles smaller than a critical size (about 8 nm) can eventually be converted into hollow iron oxide particles.[52, 53] In the sample *HNC-1* it appears that the size of the originally grown Fe-metal nanodomains (i.e. before the oxidation had commenced) was below that critical size and additionally the formation of the polycrystalline shell most likely prevented the formed Fe@Fe$_x$O$_y$

nanodomains from sustaining a hollowing process. In the case of the larger core@shell domains in the *HNC-2*, with lower degree of Fe@Fe$_x$O$_y$ coverage, the initially grown Fe-metal nanodomains were large enough to allow for the Kirkendall effect to operate and produce a void region.

Let us now elaborate on the influence of the ZnO seeds on the heterogeneous nucleation and growth of the secondary material phase to yield different sets of HNCs. For a fixed amount of Fe(CO)$_5$ injected at a given ZnO concentration in the flask, the Fe@Fe$_x$O$_y$ domain population deposited on the NRs could be controlled by using variable-size ZnO seeds developed in a mixture of ligands with different composition and regulating the Fe(CO)$_5$ decomposition temperature (Figures 1d-i, S2). Indeed, HNCs seeded with smaller ZnO NRs grown at a reduced temperature and lower amine to Zn(Ac)$_2$ molar ratio were eventually covered by a large number of small Fe@Fe$_x$O$_y$ domains (sample *HNC-1*, Figure 1d-e). On the other hand larger-volume ZnO NRs grown at a higher temperature and at a higher amine to Zn(Ac)$_2$ molar ratio supported a comparatively lower density of larger Fe@Fe$_x$O$_y$ nanoparticles (sample *HNC-2*, Figure 1g-h). Such an observed trend is opposed to what could be otherwise expected on the basis of the relative seed to Fe(CO)$_5$ proportions, realized in the syntheses of the respective cases.[1,2,35] Thus, it should be inferred that the surface features of the NRs (i.e. defects on exposed facets, degree of organic coating, nature of the capping ligands thereon), which are indeed likely to govern the heterogeneous nucleation events, varied substantially as a function of the ZnO synthesis conditions, thereby impacting on the ultimate the Fe@Fe$_x$O$_y$ domain deposition mode. The result of inducing Fe deposition in the absence of any extra surfactants added to the bulk solution suggests that the ZnO NRs became chemically accessible when the pristine capping molecules were displaced from the seed surfaces to re-establish ligand desorption-absorption equilibrium. Under such exceedingly low concentration of capping agents in the environment, any surface-preferential ligand adhesion mechanisms that could accentuate the potential facet-dependent reactivity of the NRs could hardly operate. This explains why, although the seed shape anisotropy may be expected to drive site-specific Fe deposition (e.g. onto the apexes), nevertheless differences in chemical accessibility among the various crystal facets exposed should be largely attenuated, guaranteeing a satisfactory degree of heterogeneous nucleation at the cost of topological selectivity (in fact, both the number and the locations of the Fe domains deposited on the seeds are randomly distributed).[1,2,35]

The crystallinity of the HNCs was also confirmed by XRD measurements (Figure 2). The sharp reflections of the ZnO wurtzite structure of the NRs remain unaltered in width and intensity after their incorporation into the HNCs, indicating preservation of the seed structure and geometry during the seeded-growth step (Figure S4). The additional Bragg reflections at 2θ= 44.7°, 65.0° and 82.3° in the patterns of the HNCs can be clearly indexed to the bcc structure of metallic α-Fe, providing firm evidence for the chemical nature and the crystallinity of the spherical domains grown on the ZnO NRs. We used the Scherrer equation and the (110) Bragg peak of the bcc α-Fe phase to estimate the mean diameter the Fe-metal core. The calculation suggests that the latter is about 7.9 and 20.8 nm for the two HNC samples shown in Figure 1, respectively. The difference between these values and those estimated from the HRTEM images is somewhat due to the polydisperse character of the core@shell nanodomains. In addition, the low-angle Bragg reflections at 2θ=14.0°, 16.8°, 18.4° and 21.4° which are mainly resolved for the *HNC-1* sample, can be indexed to the tetragonal γ-Fe$_2$O$_3$ and cubic Fe$_3$O$_4$ structures. This indeed confirms the presence of such Fe-based oxide phases in the partially and fully oxidized Fe–based decorating nanodomains. On the other hand, the Fe$_x$O$_y$ shell thickness could only be reliably measured through the HRTEM images and found to be ~1 nm for the *HNCs-1* and ~4-5 nm in the *HNCs-2* samples. It is worth noting that similar core-shell structures, with Fe monocrystalline core and Fe$_x$O$_y$ polycrystalline shell, were obtained also under the same synthetic conditions but without using ZnO seeds in the synthesis protocol (Figure S5a). However, the absence of an appropriate coordinating surfactant led to a higher degree of aggregation.

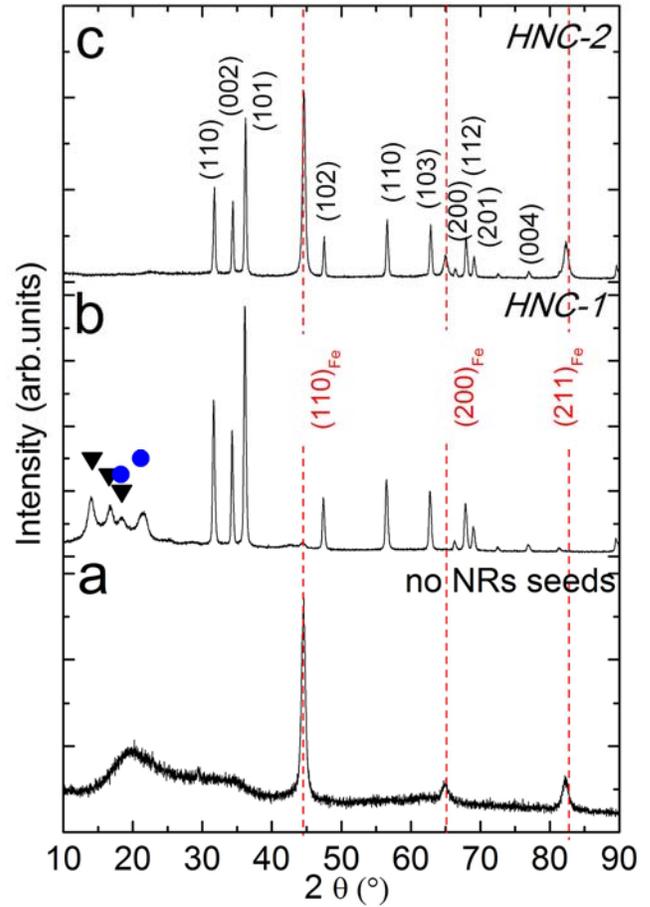

**Figure 2**. Typical powder XRD patterns of the samples without ZnO NRs seeds (a), HNCs with high (b) and low (c) degree of Fe@Fe$_x$O$_y$ domain surface coverage. The low-angle reflections (2θ< 25°) are indexed in the tetragonal γ-Fe$_2$O$_3$ (▼) (ICDD, 00-013-0458) and the cubic Fe$_3$O$_4$ (●) (ICDD, 01-079-0416); high-angle indexing based on wurtzite ZnO (ICDD, 00-036-1451)

**3.2 Surface coverage of the ZnO nanorods and optical properties.** The variable-dimension ZnO NRs were studied by PL spectroscopy. The qualitative features of the PL spectra at room temperature are comparable to those reported in the literature for anisotropic ZnO nanostructures, with somewhat different average diameter, but prepared with similar methods[54]. The nanorods were found to exhibit a pronounced near band-edge (NBE) UV emission located at 3.211 eV, for the seeds of the *HNCs-1* (Figure 3a, upper curve) and at 3.249 eV for ZnO seeds used in the growth of the *HNC-2* sample (Figure 3b, upper curve). As the ZnO NR dimensions are much larger than the material's exciton Bohr radius, $\alpha_B \sim$ 2.34 nm, the difference in the NBE between the two batches is less likely to be due to quantum confinement effects. In nanostructures of relatively larger dimensions another reason for the modified energy of the NBE emission needs to be considered. The somewhat different conditions of the colloidal synthesis for the two ZnO seeds corroborate a possible subtle variation in their surface structural details (e.g. level of defects on exposed facets). Earlier studies have shown that the defect density can be larger on the surface than in the bulk of ZnO.[55] In that respect the PL-measured variations in the position of the NBE peak in the NRs could be due to different levels of impurities present. In nanostructures with different surface to volume ratios, like here, this would influence the electronic states in the band gap and could incur small variations in the NBE transitions. In effect it can also point to a variable carrier concentration amongst the two batches of the NRs. As that, the lower-energy emissions in the present case (with respect to the expected ~3.36 eV values for highly crystalline defect-free ZnO crystals) may suggest higher carrier concentrations for the NR seeds. In addition, a broader and much weaker band emission, related to such deep level defects (DL; green luminescence) was also detected in the visible spectral region (~2.4 meV) for all samples.

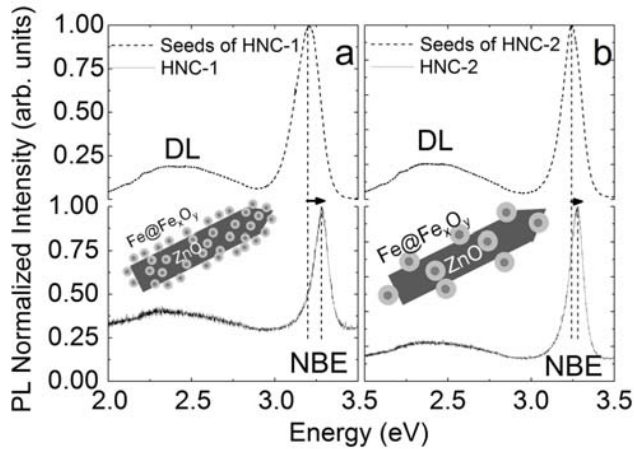

**Figure 3.** PL at 300 K of the *HNC-1* (a) and *HNC-2* (b) samples compared to the corresponding ZnO NRs seeds (top spectra). The vertical dash line shows the blue shift of the NBE emission due to the Fe@Fe$_x$O$_y$ coverage attained over ZnO. Each spectrum was normalized to the associated NBE UV emission max intensity.

On the other hand the HNC samples were found to exhibit a NBE UV emission located at about 3.286 eV for the *HNC-1* (Figure 3a) and 3.277 eV for the *HNC-2* (Figure 3b). An important observation though, concerned with the HNCs, is that the NBE emission is strongly shifted to the blue spectral region when it is compared to that of the parent NR seeds (Figure 3). The NBE spectral shift upon coverage is $\Delta_{NBE}$= +75 meV for the *HNC-1*, while it is moved by $\Delta_{NBE}$= +28 meV for the *HNC-2* sample. We infer that the observed modifications in the optical properties of the ZnO should be a consequence of the coverage of its surface by the Fe@Fe$_x$O$_y$ nanodomains. Earlier studies on 10 nm core@shell FePt@ZnO HNCs have shown that the highly deformed lattice of the semiconductor shell (due to a compression along the c-axis) induced by interfacial connection with FePt core could modify the ZnO optical response in a similar way.[15] In another account, in relatively larger ZnO nanostructures (diameter of ~100 nm), PL experiments have illustrated a pronounced pressure-dependent shift of the UV emission towards higher energy (pressure coefficient ~23-24 meV/GPa), as a consequence the ZnO band-gap increase.[56, 57] Evidence in support of these cannot be directly drawn from our combined HRTEM and XRD structural studies. However, at this stage we cannot waive out the possibility that a strongly piezoelectric field is induced in the underlying ZnO lattice by (compressive or tensile) strain generated due to the extended interface shared between the lattice mismatched nanorods and Fe@Fe$_x$O$_y$ nanocrystals decorating them (Figures 1e, h). The larger blue shift in the *HNC-1* sample may originate from the enhanced interfacial interactions as a result of the more extended surface coverage of the ZnO NRs in this case.

Furthermore, the sharp and intense NBE UV emission of these ZnO nanomaterials can be attributed to exciton recombination, in agreement with the literature.[58] Besides, the deep level emission is usually related to oxygen vacancies, surface states and other structural defects. In the present work the DL in the HNCs was found to be of relatively lower intensity as compared to that in the associated seeds and the ratio of the NBE over the DL band intensities, $I_{NBE}/I_{DL}$, was enhanced when the ZnO NRs were covered by the Fe-based NCs. To a first approximation, the increased ratio in the case of HNCs could indicate that either the Fe@Fe$_x$O$_y$ domain coverage or the heating of the ZnO NRs at high temperature during the step of the heterogeneous nucleation of the Fe-based nanodomains led to surface defect passivation and/or annealing, thus suppressing the related green emission band. In this respect, the relative increased intensity of the NBE compared to the DL emission indicates good crystallinity and lower defect density for the HNCs synthesized by the proposed seeded-growth approach.

**3.3 Magnetic exchange coupling mediated by frozen interfacial spins.** In Figure 4 we show the room-temperature hysteresis loops for the same representative samples addressed in Figure 2. The presence of metallic Fe in the magnetic domains rendered both samples of the HNCs ferromagnetic, with coercive fields (H$_c$) of 336 and 369 Oe and saturation magnetizations (M$_s$) of 2.3 and 61.0 emu/gr for the *HNC-1* and *HNC-2*, respectively. These characteristic values are lower compared to those exhibited by the corresponding Fe-based materials in their bulk form, where M$_S$ reaches 74 emu/g for $\gamma$-Fe$_2$O$_3$, 84 emu/g for Fe$_3$O$_4$ and 219 emu/g for Fe-metal micro-powders. The larger Fe core in the *HNC-2* gave a relatively larger coercive field and a significantly enhanced saturation magnetization as compared to those measured for *HNC-1*, which indeed involved a smaller-diameter Fe cores. To evaluate the effect of the diamagnetic contribution of the ZnO section on the HNC magnetic properties, we extended our study to the parent ZnO NRs seeds (inset, Figure 4). Their diamagnetism is found to be negligible with respect to the magnetization values obtained for the HNCs under the same experimental conditions. Furthermore, similar core@shell Fe@Fe$_x$O$_y$ structures were prepared under the same conditions in the absence of the ZnO NRs seeds. This material displayed also ferromagnetic like behaviour at room temperature, with H$_c$= 307 Oe and M$_s$= 13.7 emu/gr (Figures 4 and S5b).

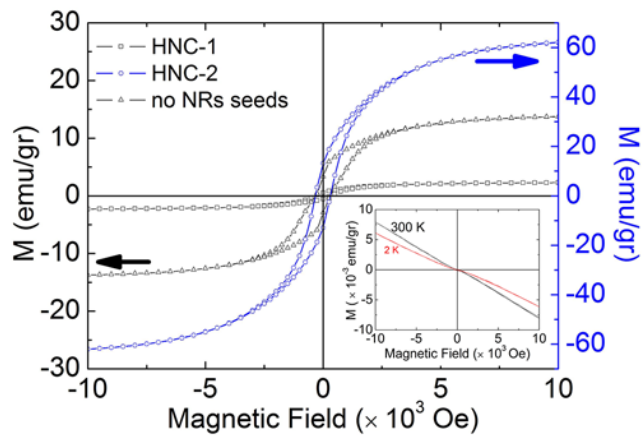

**Figure 4**. Hysteresis loops at room temperature of the samples synthesized without the ZnO NR seeds (triangles), HNCs with high (squares) and low degree of Fe@Fe$_x$O$_y$ domain coverage (circles). Inset: the hysteresis loops of the bare ZnO NRs at 300 K and 2 K.

It is worth noting that, although the magnetic properties of the *HNC-1* were unchanged with the time, those concerned with the *HNC-2* were sensitive to the structural evolution of the Fe@Fe$_x$O$_y$ domains due to the Kirkendall effect. We observed that the coercive field of the *HNC-2* at room-temperature was considerably reduced after a few months of the powder sample being stored at ambient conditions. The measured hysteresis loop indicates the drastic reduction of the $H_c$ from 369 Oe in the as-prepared samples to 62 Oe, after 2.5 months (Figure 5). In conjunction with the HRTEM studies, we suggest that a nearly complete separation of the Fe core from the Fe$_x$O$_y$ shell takes place, with no detectable magnetization changes beyond this time span.

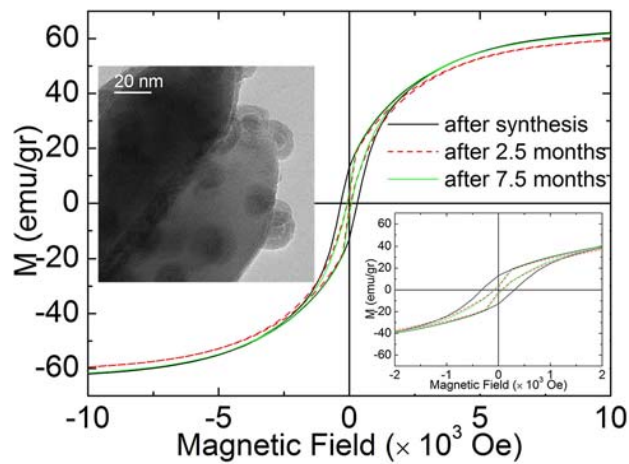

**Figure 5.** Hysteresis loops for the HNC-2 sample at 300 K, taken at different periods of time after the sample preparation. Inset: a HRTEM image of this sample after 7.5 months (left), and the low-field region of M(H), with marked reduction of $H_c$ over time (right).

The previous sections corroborate that the room-temperature ferromagnetism of the HNCs is due to the binary domains made up of a ferromagnetic (FM) core interfaced with a ferrimagnetic (FiM) shell. In this case two magnetically unlike phases exist across a common interface that can potentially mediate strong magnetic exchange-coupling mechanisms[59]. Unidirectional coupling between FM and antiferromagnetic (AFM) or FiM (like in the present case) layers provides, in general, a route for the technologically-important stabilization of the magnetization in nanoscale systems via the so-called exchange-bias (EB) effect.[60] This is a measurable quantity and is typically reflected in the detection of a horizontal shift of the material's hysteresis loop by an EB field, upon cooling in an applied magnetic field.[61]

To recognize and evaluate possible exchange coupling effects in our HNCs, arising from the magnetic core@shell geometry of the decorating domains, we have studied their magnetization hysteresis curves, M(H), at 5 K, under ZFC and FC conditions, as described in the experimental section. Under FC conditions, we found a negative loop shift for *HNC-1* sample, decorated with the small core@shell domains, indicating that magnetic exchange-coupling was indeed established due to the common interface between the FM core and the FiM shell (Figure 6a). The EB field, $H_{eb}$, was 208.8 Oe, a value similar to that measured for Fe/Fe oxide core@shell nanocubes[62] and even for Fe nanoparticles embedded within an iron oxide matrix.[63, 64] Its value was estimated from the measured hysteresis loop curves, through the equation $H_{eb}= |H_1^{FC}+H_2^{FC}|/2$, where $H_1$ and $H_2$ are the right and left intercepts of the curve along the field axis. The origin of the EB has earlier been attributed to uncompensated magnetic moments of the FiM (or AFM) domain that are pinned at the interface with the FM layer.[65] In support to this argument came the dependence of the $H_{eb}$ on the cooling field strength. We found that its magnitude (at 5 K) grew with increasing applied field, while it became somewhat reduced and saturated for cooling fields higher than 10 kOe (inset, Figure 6a). The quick rise of the $H_{eb}$ reflected the presence of frozen interfacial spins in the FiM material, which progressively became aligned (in larger numbers) to a certain degree along the applied field direction. Earlier studies on diverse materials, for example, on granular systems composed of Fe nanoparticles embedded in an iron oxide matrix[63] or in others concerned with monodisperse, isolated Fe@Fe$_3$O$_4$ core@shell nanoparticles[66], have also offered similar explanations. Furthermore, at higher cooling fields values (>10 kOe), the Zeeman coupling may prevail over the effect due to exchange coupling, rendering the spins at the FiM shell less strongly coupled to the FM core.[63, 67] As a result the polycrystalline shell can generate weaker pinning to the interfacial spins in the FM layer, which in turn may lead to partly reduced $H_{eb}$ and coercivity (Figure S6). On the other hand, the *HNC-2* sample with the large core@shell domains exhibited no shift in the FC hysteresis loop ($H_{eb} \sim 0$ Oe). As the magnetic exchange coupling decays exponentially with the distance, the existence of voids intervening between the Fe core and the Fe$_x$O$_y$ shell in the corresponding Fe@Fe$_x$O$_y$ domains of the HNCs led to diminution of the pinned interfacial moments and their effect. The lack of EB in *HNC-2* sample highlights the role that the subtle structural differences may play on the nanoscale.

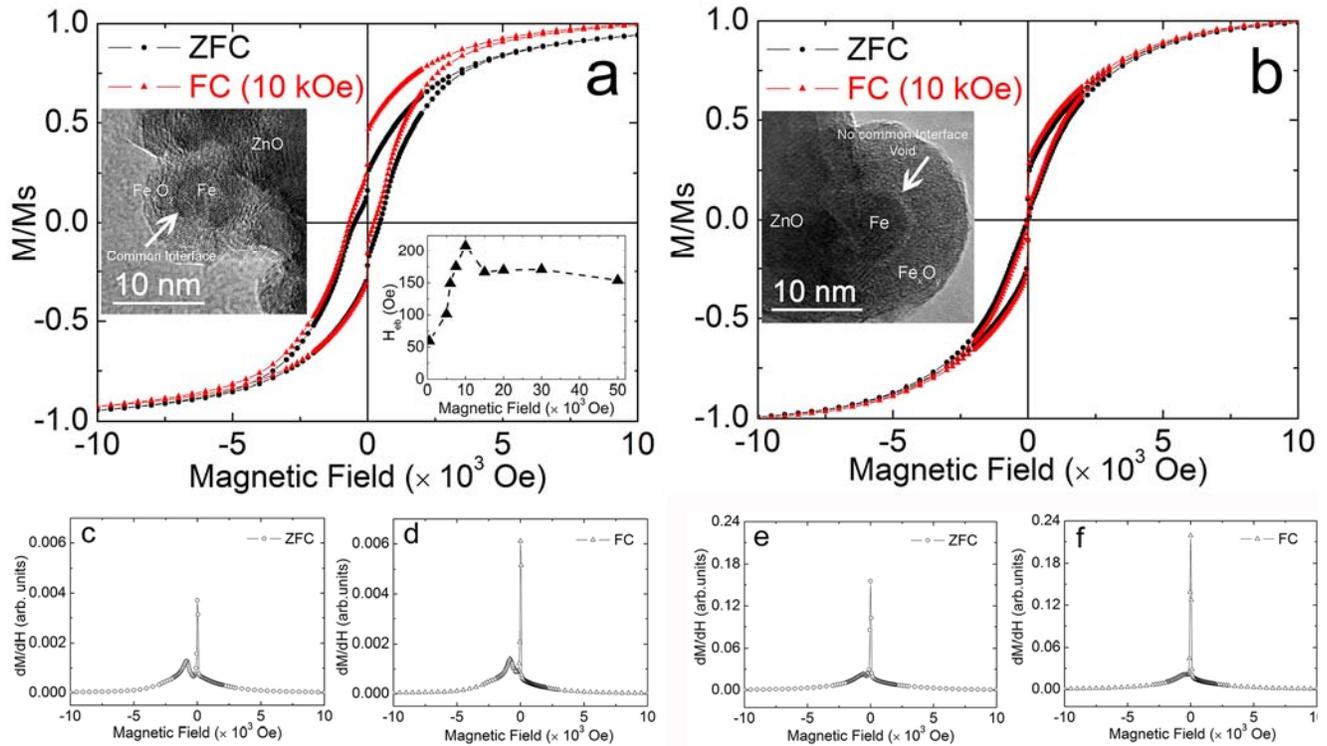

**Figure 6.** Magnetic hysteresis loops of the *HNC-1* sample, with $M_s$= 2.5 emu/g (a) and *HNC-2* sample, with $M_s$= 61.2 emu/g (b) at 5 K, under FC (triangles) and ZFC conditions (circles). (c-f) dM/dH plots for the *HNC-1* and *HNC-2* samples during the negative sweep direction (from +10 to -10 kOe) under ZFC (circles) and FC (triangles) conditions. Insets: (a) $H_{eb}$ as a function of the cooling field strength. (a, b) HRTEM images of the Fe-based domain and its evolution.

The net magnetic moment generated by the small percentage of the frozen interfacial spins can be calculated by the formula $M_f = \Delta M/2 = (M^+ + M^-)/2$,[68] where $M^+$ and $M^-$ are the saturation magnetization in the positive and negative field directions (± 10 kOe) under FC conditions. Based on that, the hysteresis loops at 5 K gave an $M_f$= 3.72×10$^{-2}$ emu/gr for *HNC-1*. The relative magnitude of the frozen spins' magnetic moment was 3.7% of the total value. In addition to the horizontal shift, the normalized magnetization $(M/M_s)$ was found to be asymmetric with respect to the origin. Such a vertical shift of the hysteresis loop can also be attributed to the frozen spins of the shell in FM/FiM core@shell nanostructures.[59] The good crystallinity of the $Fe_xO_y$ shell allows for a large number of pinned interfacial spins and a sizeable exchange bias effect. To verify this we modified the synthetic protocol so that the Fe-based domains were developed with an amorphous shell (Figure S3). We found then that the magnetic moment of the frozen spins was very much reduced (~ 1 % of the total value) and in turn a negligible exchange bias field, $H_{eb}$ = 3.5 Oe, was measured (Figure S3d).

*Low-field spin reorientation.* In addition to the horizontal shift of the loops, we also observed a discontinuous, step-like variation of the M(H) at low fields (-50≤ H≤ 50 Oe), under both ZFC and FC conditions. The derivative plots of the isothermal magnetization, dM/dH (Figure 6c-f), nicely verify this effect for both our samples and suggest its origin to the shell. We quantified the low-field jump by means of the overall magnetic moment changes, $\Delta M/M^+$, as done in previous studies of isolated core@shell nanoparticles.[68] $\Delta M$ is the reduction of the magnetization at low fields and $M^+$ is the saturation magnetization in the positive field sweep direction for the hysteresis loop (+10 kOe). Under FC conditions the jump accounts for 23.2 % of the overall magnetization (14.9 % at ZFC) for the *HNC-1* sample, while for the *HNC-2* is 26.2 % (24.8 % for the ZFC). At low fields the Zeeman energy, which couples the moments to the external applied field, is dominated by the magnetocrystalline anisotropy energy. As a result, the latter favors alignment of the spins along the easy-axis direction for each one of the polycrystalline shell's domains (~3-5 nm size; refer to HRTEM images in Figures 6 and S2). This low-field reorientation of the moments appears to be sensitive to nanoscale structural parameters of the Fe@$Fe_xO_y$ domains, including, the thickness of the ferrimagnetic shell and the volume ratio between the core ($V_{core}$) and the shell ($V_{shell}$).[66, 68] For this reason we calculated the effective volume ratio, $V_{shell}/V_{core}$, as a means to measure its impact. On the basis of the ratio being 0.42 for the domains in *HNC-1* and 0.69 for *HNC-2*, it is inferred that the relative thinner shell of the former justifies the somewhat smaller demagnetization jump ($\Delta M$) and the stronger enhancement under FC cooling (+8.3 % for the *HNC-1*, against +1.4 % for the *HNC-2*).

*Temperature-dependent magnetization.* In Figure 7 we show the ZFC-FC magnetization curves (T= 5-300 K and H= 10 kOe) for the same representative HNCs samples. Even after such a strong magnetic field is applied, a notable "bifurcation" ($T_B$) in M(T) is identified at low temperatures (Figure 7). This effect provides additional support to the role of frozen spins in the two different HNCs samples. While the magnetic moment variation, $M_{FC}-M_{ZFC}/M_{FC}$ at 5 K, is 5.7% and 0.7% for the *HNC-1* and *HNC-2*, respectively, the $T_B$ is lower for the former (~30 K) and higher (~50 K) for the latter. Qualitatively, the magnetic moment variation under this high magnetic field can be attributed to the incomplete alignment of the frozen spins at the interface between the core and the shell, as well as those spins lying at the surface of the shell. Magnetic moments of this origin are pinned at random directions when the sample is cooled without an applying magnetic field, but they are aligned to a certain extent, under 10 kOe, along the field direction; a hysteresis in M(T) ZFC-FC curves is then measured. Above $T_B$ the thermal energy overcomes the magnetic anisotropy energy and the spins can then be aligned in the field direction. The observed relative difference of the M(T) magnitude at 5 K, between ZFC and FC protocols in the two samples, is likely due to the additional, uncompensated magnetic moments of the FiM domain that are pinned at the interface with the FM layer in the core@shell domains of the *HNC-1* sample and their absence in the case of the *HNC-2*. A plausible source for the increased $T_B$ in the latter can be the larger volume ratio effect as compared to that for the *HNC-1*. Such a behavior has been observed in similar nanoparticle topologies with comparable size cores.[69]

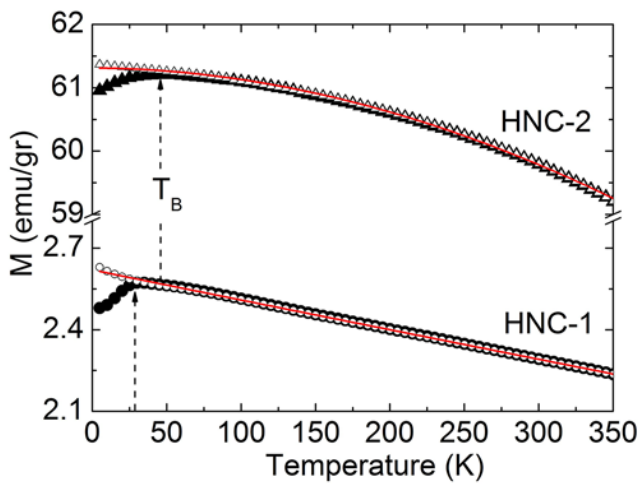

**Figure 7**. ZFC (full symbols) and FC (open symbols) temperature dependent magnetization curves, M(T), for the *HNC-1* (circles) and *HNC-2* sample (triangles) taken at H= 10 kOe. The solid, red lines represent the fit to the data based on the Bloch's law for spin-wave excitations.

Interestingly, at $T>T_B$ the monotonic thermal decrease of the magnetization (Figure 7) was found to obey a power-law, very closely resembling that of the conventional $T^{3/2}$-Bloch's behavior for spin-wave excitations in bulk ferromagnetic materials. The FC curves were fitted to the equation $M(T) = M_S(0)[1- B_0 T^n]$, where $B_0$ is the Bloch constant and n is the corresponding exponent. The derived exponents were 0.99±0.01 ($B_0= 1.1±0.1 \times 10^{-3}$ K$^{-n}$) for the *HNC-1* and 1.94±0.02 ($B_0= 2.3±0.2 \times 10^{-5}$ K$^{-n}$) for the HNC-2. We note that, although the shell thickness may play a role here, $B_0$ is decreased with increasing core size, as it is expected when the physical dimensions of a structure tend to the bulk ($B_0^{bulk}= 3.3 \times 10^{-6}$ K$^{-3/2}$).[70] Although earlier studies have shown that the $T^{3/2}$ law is followed by a number of nanostructured systems, deviations from the ideal value could suggest modifications of the local magnetic structure due to extra sources of local magnetic anisotropy. The reduced coordination at surfaces and interfaces in the nanoscale may be a reason behind. Such magnetic moment topologies can be more susceptible to thermal excitations with altered from the bulk behavior. For example, it has been postulated that frustrated antiferromagnetic interactions due to partial surface oxidation in FePt nanoparticles attained exponents, n> 1.5,[71] whereas possible magnetic exchange coupling effects in core@shell architectures of Fe nanoparticles led to reduced values, n< 1.5. [70, 72]

## CONCLUSIONS

We have presented a seeded-growth colloidal protocol that affords a novel nanoheteroarchitecture. Our approach allows for modification of its structural characteristics so that it can display room temperature magneto-optical response. The as-formed hybrid nanocrystal structure is comprised of quasi-spherical core@shell Fe@Fe$_x$O$_y$ domains that are assembled on bullet-like ZnO nanorods, with no facet-preferential attachment. As a consequence the system exhibits collective physical properties. Purposeful modification of the alkyl amine surfactant to Zn-precursor molar ratio, in conjunction with regulation of the temperature at which the ZnO NR seeds are grown allows control over the number of the Fe-based nucleation sites, as well as the development of larger in size Fe@Fe$_x$O$_y$ domains. The polycrystalline shell in the latter, with its randomly oriented crystallites, is shown to prevent the Fe-domains from complete oxidation. We have demonstrated that structural differences at the nanometer scale, between as-synthesized samples, have strong impact on their physical behavior. The magnetic nanocrystals can be tailored to grow with a common interface between the ferromagnetic Fe core and the ferrimagnetic Fe$_x$O$_y$ shell, while size-dependent sacrificial conversion of the core due to the Kirkendall effect is further utilized as a means to structurally decouple the two unlike magnetic phases. In the former case uncompensated frozen interfacial spins mediate the exchange-bias effect in the hybrid, while in the hollowed nanostructure the phenomenon is found to be suppressed. Furthermore, our optical studies suggest that the surface coverage of the semiconducting ZnO NRs with the Fe@Fe$_x$O$_y$ nanocrystals influences ZnO photoluminescence. The latter experiments are especially enlightening in that the near band edge UV-emission of ZnO is found to be strongly shifted to the blue spectral region. The results cannot exclude the possibility that the interfacial connectivity, found between the NRs and the Fe-based domains, can generate a strained ZnO lattice which in turn influences its band-gap and the associated optical response. The afore-mentioned bifunctional hybrid nanocrystals have thus shown appreciable fluorescence and exhibit ferromagnetic-like behaviour at room temperature, not otherwise achievable with the corresponding physical mixtures of unbound ZnO and Fe@Fe$_x$O$_y$ core@shell nanoparticles. This new nanomaterial, with its collective properties, is therefore likely to find applications in broader fields including visible-light driven photocatalysis, in biomedice for cell labeling and separation, or even in spintronics.

## ASSOCIATED CONTENT

Supporting Information. TEM images, size distributions of the nanorod seeds; HRTEM images of the Fe@Fe$_x$O$_y$ core-shell NCs decorating the HNC-1 and HNC-2 ; TEM images, powder XRD pattern and hysterisis loops at5 K (under ZFC and FC conditions) and 300 K for HNCs covered by an amorphous $Fe_xO_y$ shell; Comparison of the XRD pattern of the nanorod seeds with those of the HNC-1 and HNC-2; HRTEM image, FFT and hysteresis loops at 5 and 300 K of Fe@$Fe_xO_y$ core-shell NCs synthesized in the absence of ZnO seeds; Cooling-field strength dependence of $H_{eb}$ and the $H_C$ for the HNC-1 sample. This material is available free of charge via the Internet at http://pubs.acs.org.

## AUTHOR INFORMATION

### Corresponding Author

*Corresponding author at: Institute of Electronic Structure and Laser, Foundation for Research and Technology- Hellas, P.O. Box 1385, Vassilika Vouton, 71110 Heraklion, Greece.
Tel.:+302810391344
fax:+302810391305.
E-mail address: lappas@iesl.forth.gr

### Author Contributions

The manuscript was written through contributions of all authors.

### Funding Sources

This work was supported by the European Commission through the Marie-Curie Transfer of Knowledge program NANOTAIL (Grant no. MTKD-CT-2006-042459).

## ACKNOWLEDGMENT

This work was supported by the European Commission through the Marie-Curie Transfer of Knowledge program NANOTAIL (Grant no. MTKD-CT-2006-042459).

## ABBREVIATIONS

HNCs, Hybrid NanoCrystals; NRs, NanoRods; NCs, nanocrystals; ODE, octadecene; HRTEM, high resolution transmission electron microscope; CCD, charged coupled device; XRD, X-ray diffraction; SQUID, Superconducting Quantum Interference Device, FFT; Fast Fourier Transform; ZFC, zero field cooled; FC, field cooled; EB, exchange-bias; $M_s$, saturation magnetization; $H_{eb}$, exchange-bias field; $H_c$, coercive field; FM, ferromagnetic; FiM, ferrimagnetic; AFM, antiferromagnetic; PL, photoluminescence; UV, ultraviolet; NBE, near band edge; DL, deep level

# Colloidal Anisotropic ZnO- Fe@Fe$_x$O$_y$ Nanoarchitectures with Interface-Mediated Exchange Bias and Band-Edge Ultraviolet Fluorescence

Supporting Information

Athanasia Kostopoulou,[1,2] Franck Thétiot,[1] Ioannis Tsiaoussis,[3] Maria Androulidaki,[1] P. Davide Cozzoli,[4,5] and Alexandros Lappas[1,]*

[1]Institute of Electronic Structure and Laser, Foundation for Research and Technology–Hellas, Vassilika Vouton, 71110 Heraklion, Greece

[2]Department of Chemistry, University of Crete, Voutes, 71003 Heraklion, Greece

[3]Department of Physics, Aristotle University of Thessaloniki, 54124 Thessaloniki, Greece

[4]National Nanotechnology Laboratory (NNL), Istituto Nanoscienze- CNR, c/o Distretto Tecnologico ISUFI, via per Arnesano km 5, 73100 Lecce, Italy.

[5]Dipartimento di Matematica e Fisica "E. De Giorgi", Università del Salento, via per Arnesano, 73100 Lecce, Italy.

E-mail: lappas@iesl.forth.gr

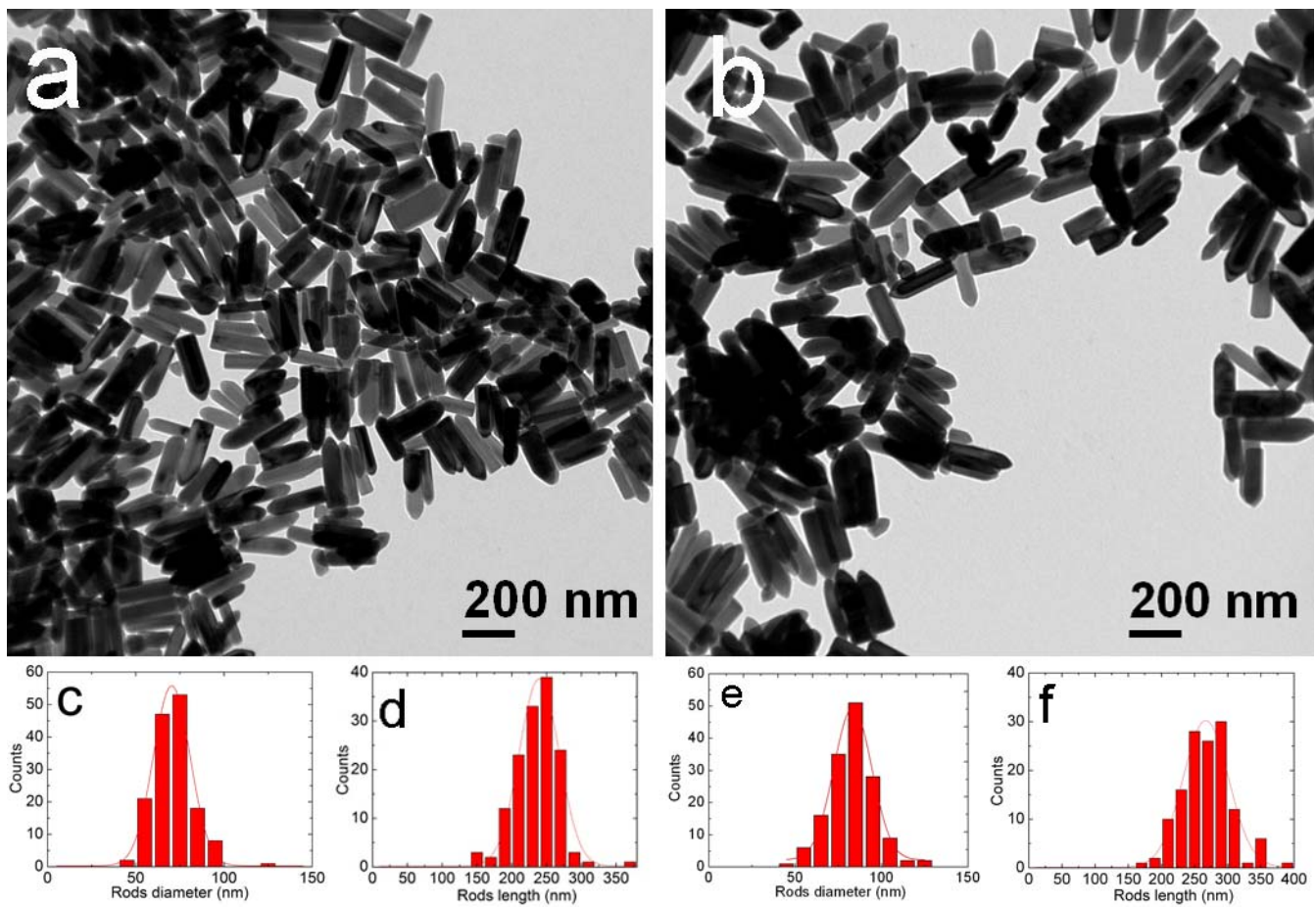

**Figure S1.** Examples of bullet-shaped ZnO NRs grown in hot alkylamines. Low magnification TEM images of the ZnO NR seeds prepared in HDA at 240 °C (a) and in a bi-surfactant HDA:OLAM mixture at 280 °C (b). Size distribution of the length, L (c, e) and the diameter, D (d, f) for the NRs grown in each case, (a)-(b).

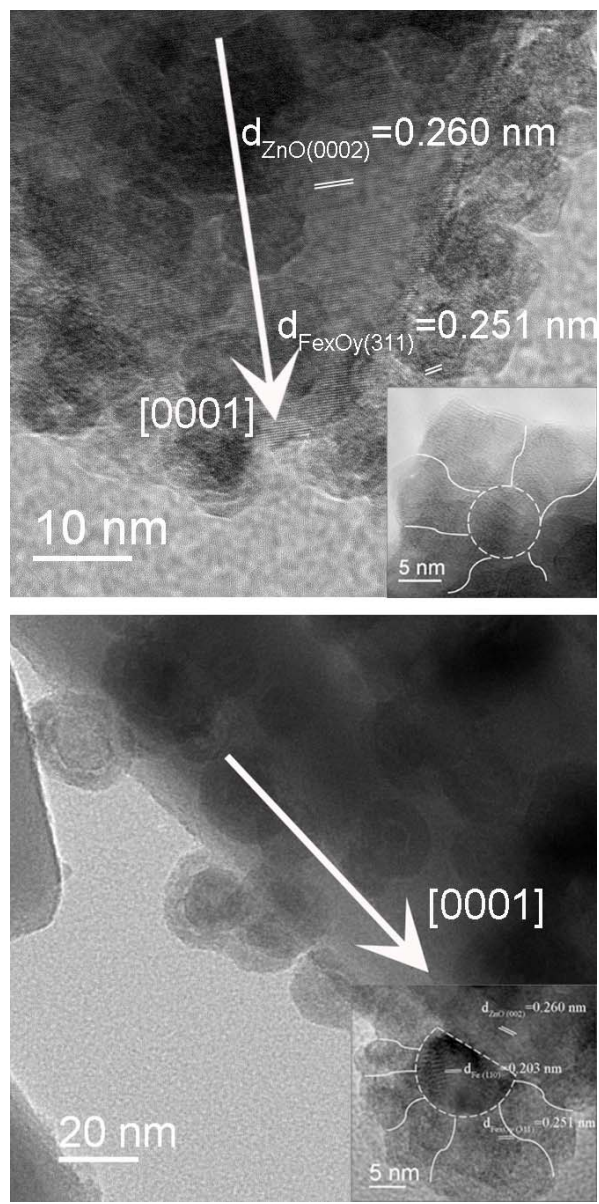

**Figure S2** HRTEM images of the HNCs with high (top) and low degree of ZnO seed surface coverage (bottom), pointing to the structural details of the individual decorating Fe@Fe$_x$O$_y$ core@shell domains grown in the two cases (insets).

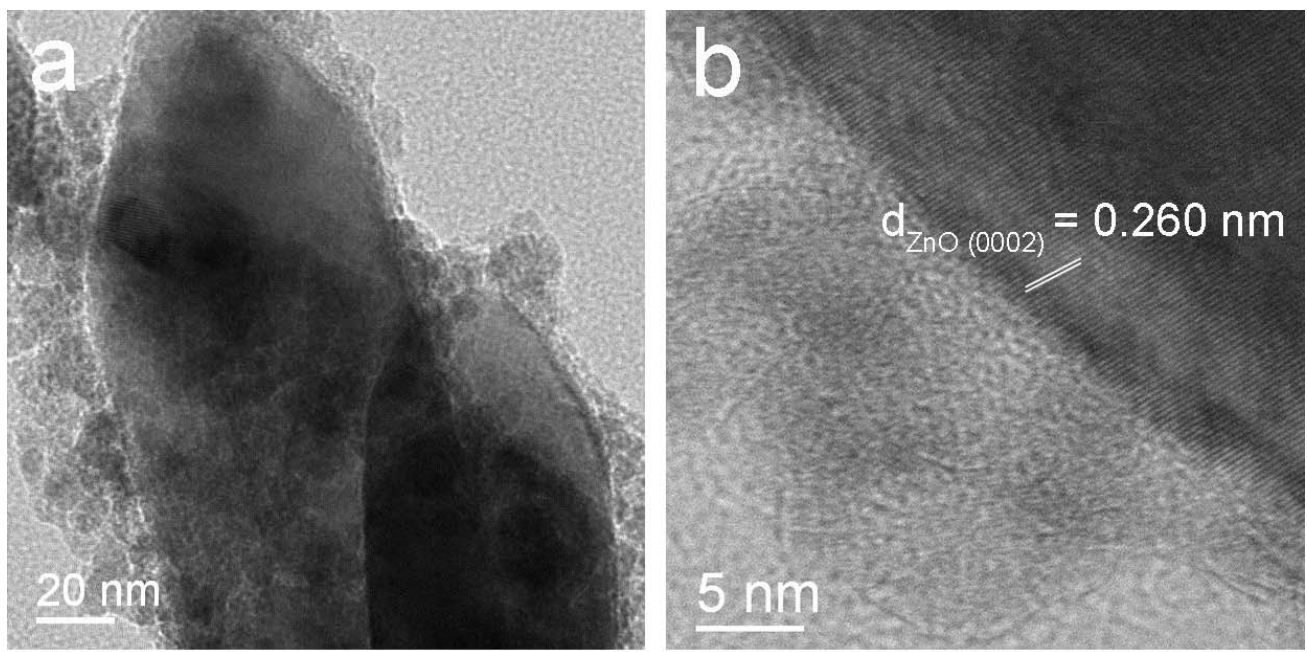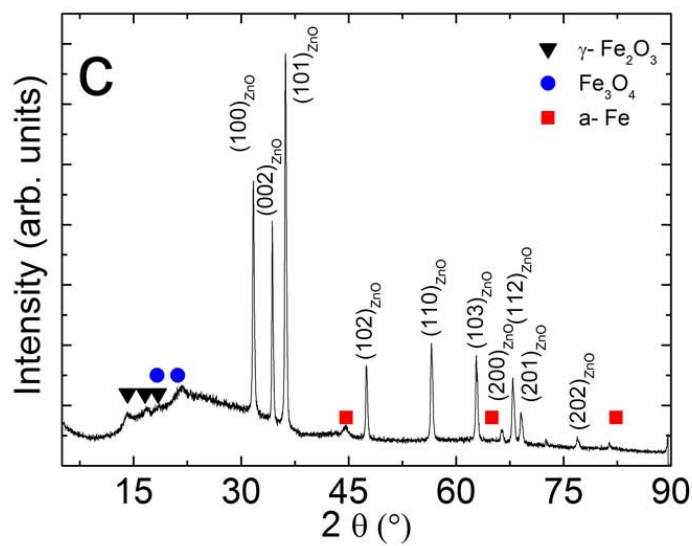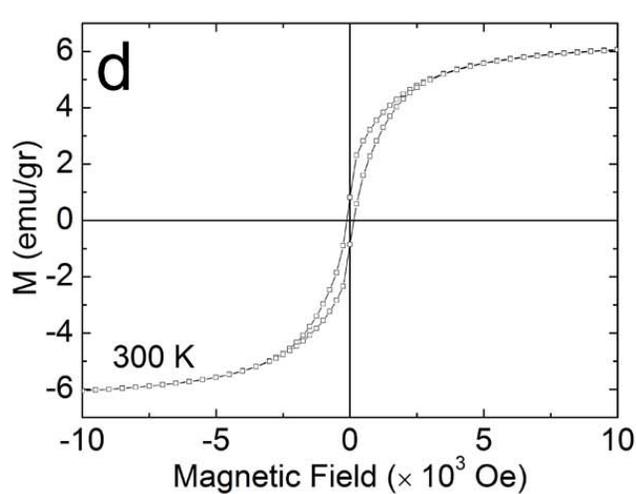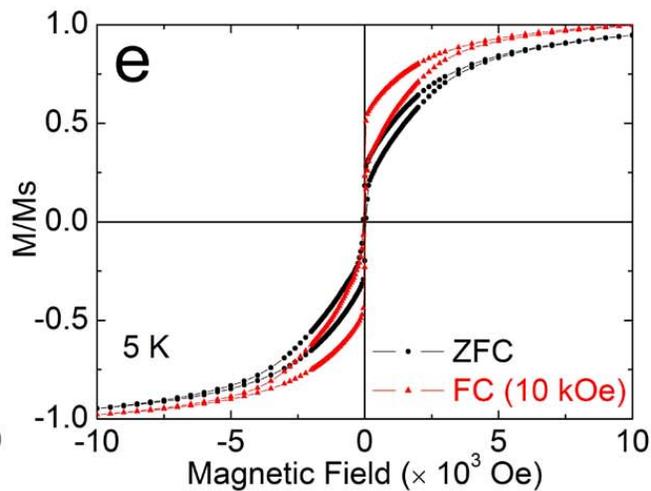

**Figure S3.** Low magnification(a), high resolution (b) TEM images, XRD pattern(c), hysteresis loop at 300 K (d) and hysteresis loop at 5 K (e) under ZFC (circles) and FC conditions (triangles) of the HNCs prepared at 215 °C.

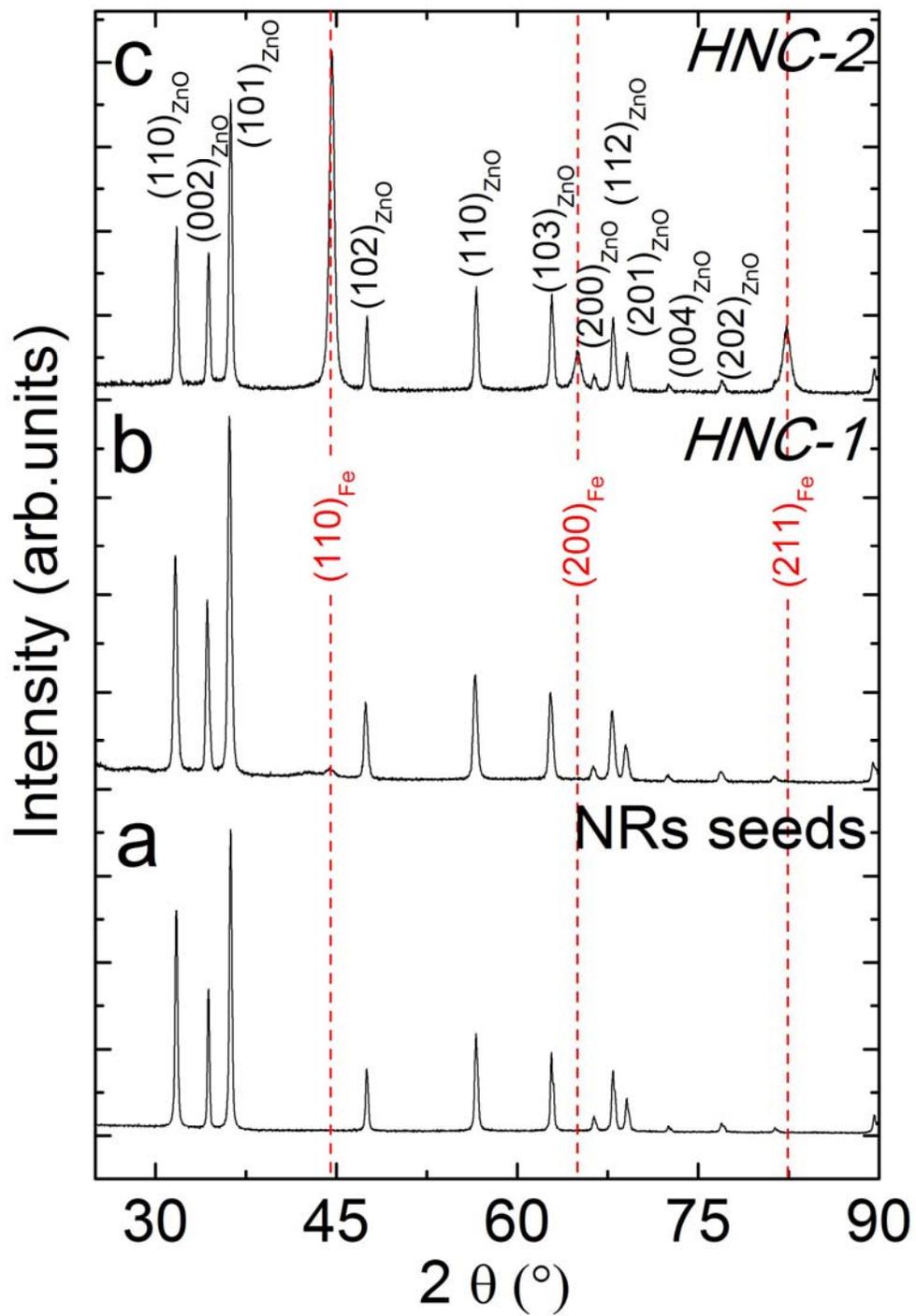

**Figure S4.** Typical powder XRD patterns of the samples corresponding to the ZnO NRs seeds (a), HNCs with high (b) and low (c) degree of Fe@Fe$_x$O$_y$ domain surface coverage.

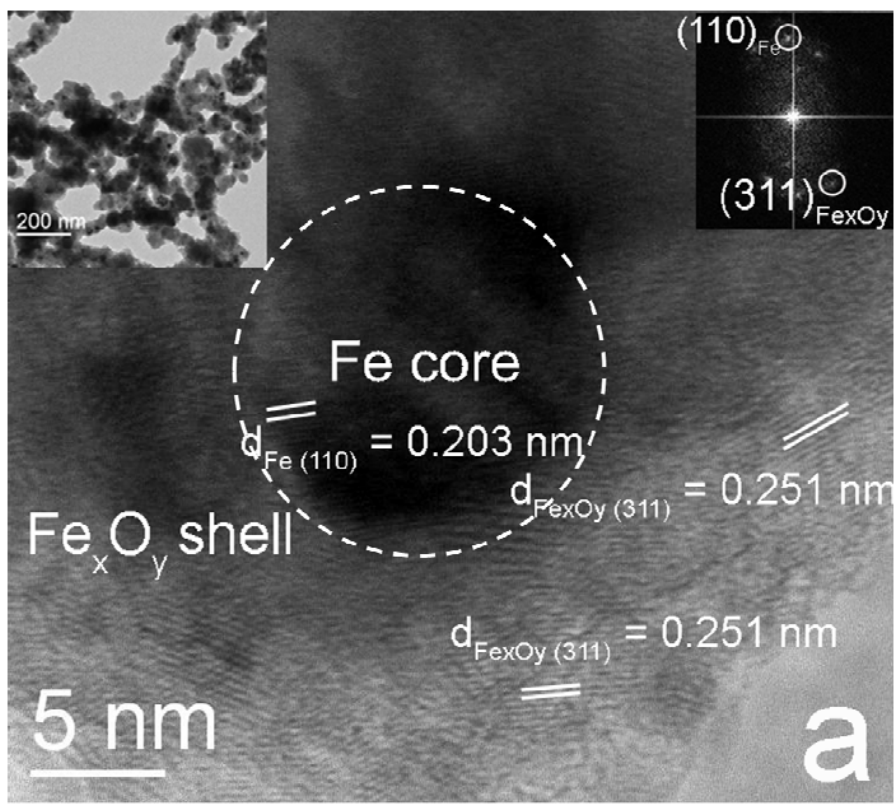

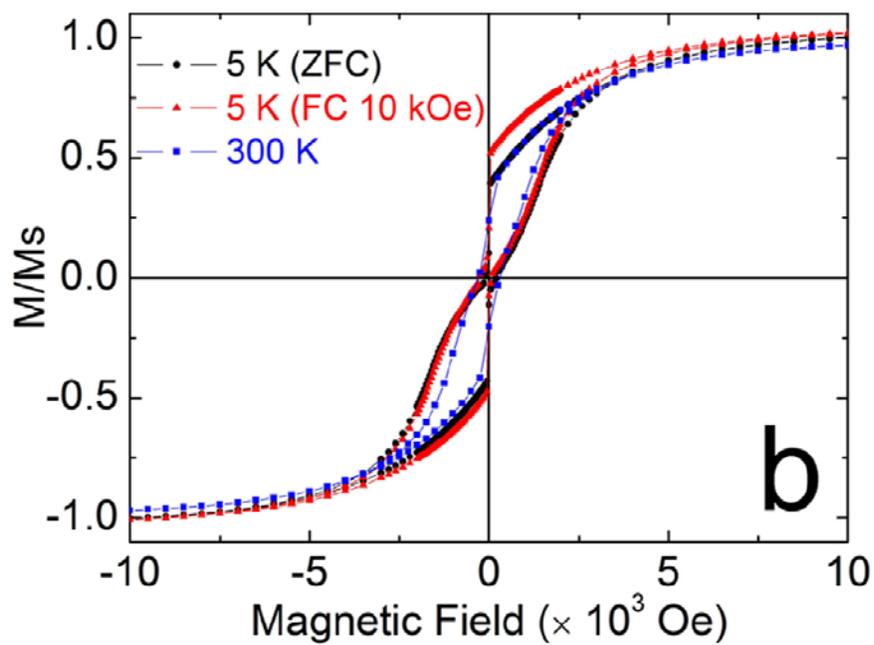

**Figure S3.** High resolution TEM image (a) and the hysteresis loop at 5 K (b) under ZFC (circles) and FC (triangles) conditions of the core@shell Fe@Fe$_x$O$_y$ NCs prepared without ZnO NRs seeds. Insets in the panel a: A low magnification TEM image and the corresponding FFT pattern of the HRTEM image.

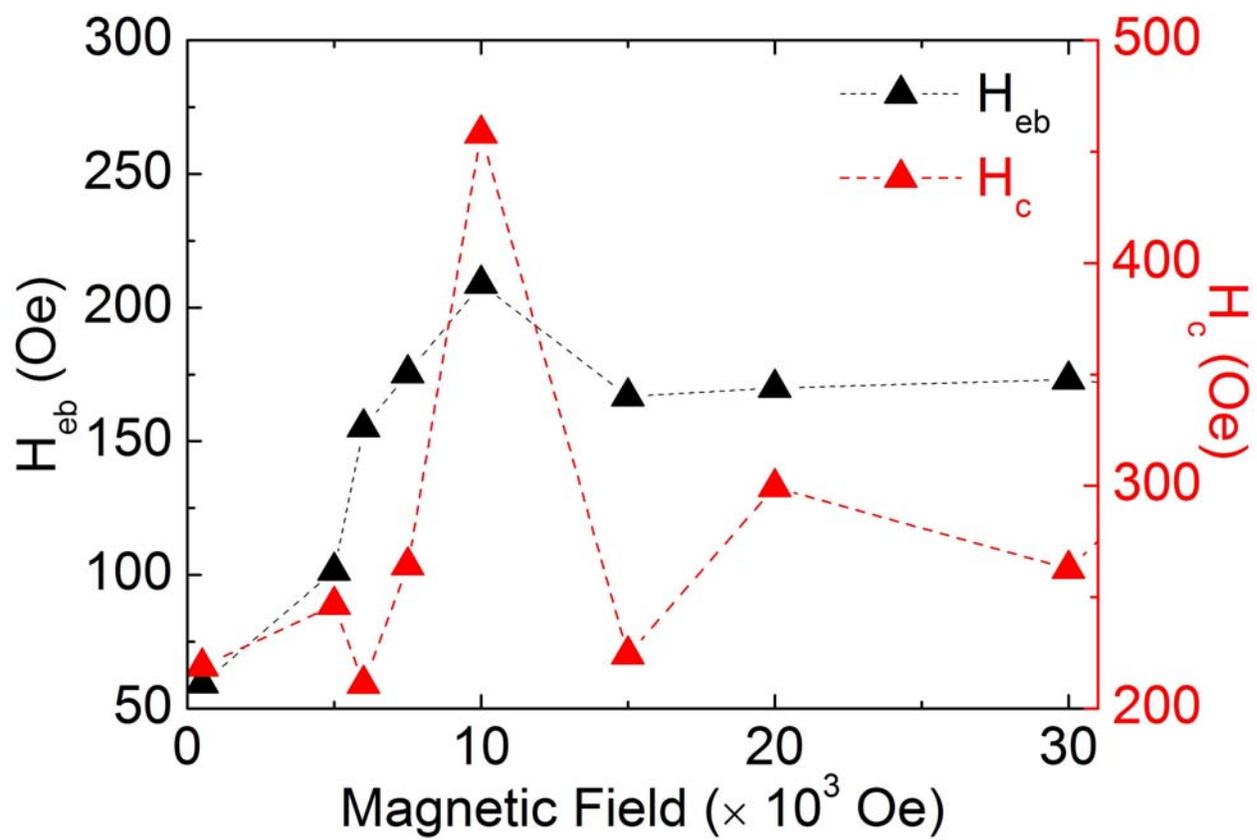

**Figure S4.** Cooling-field dependence of the exchange-bias ($H_{eb}$) field and the corresponding coercivity ($H_c$) for the *HNC-1* sample.